\documentclass[a4paper,11pt]{article}
\usepackage{pos}

\title{Coherent deeply virtual Compton scattering on helium-4 beyond the leading twist approximation}

\author[a,b]{V.~Mart\'inez-Fern\'andez}
\affiliation[a]{Universit\'e Paris-Saclay - CEA - IRFU, 91191 Gif-sur-Yvette, France}
\affiliation[b]{Center for Frontiers in Nuclear Science, Stony Brook University, Stony Brook, NY 11794, USA}

\author[c]{B.~Pire}
\affiliation[c]{Centre de Physique Th\'eorique, CNRS, École polytechnique, I.P. Paris, 91128 Palaiseau, France  }

\author[d]{P.~Sznajder}
\affiliation[d]{National Centre for Nuclear Research (NCBJ), Pasteura 7, 02-093 Warsaw, Poland}

\author*[d]{J.~Wagner}

\emailAdd{jakub.wagner@ncbj.gov.pl}

\abstract{
Coherent hard exclusive reactions on light nuclei offer access to their quark and gluon structure and provide a framework for three-dimensional nuclear tomography. We analyze deeply virtual Compton scattering on a helium-4 target, including leading-twist contributions as well as kinematical twist-3 and twist-4 effects. We present numerical estimates of cross sections and asymmetries for kinematic regimes relevant to JLab experiments. We emphasize the importance of next to leading order (NLO) and kinematical twist 3 and 4 contributions to interpret data in the JLab kinematics. We deduce a first tomographic image of quarks and gluons in the helium-4 nucleus.}

\FullConference{The 33rd International Workshop on Deep Inelastic Scattering and Related Subjects (DIS2026)\\
4 - 8 May 2026\\
Bologna, Italy\\}

\begin{document}
\maketitle
\section{Introduction}
Deeply virtual Compton scattering (DVCS) is the golden channel to access generalized parton distributions (GPDs)~\cite{Muller:1994ses,Ji:1996nm,Radyushkin:1998bz, Diehl:2003ny, Belitsky:2005qn} and perform a quark and gluon tomography of hadrons~\cite{Burkardt:2002hr,Ralston:2001xs,Diehl:2002he}. While the nucleon case is the most studied target, the quark and gluon structure of light nuclei~\cite{Berger:2001zb, Scopetta:2004kj,Kirchner:2003wt,Cano:2003ju, Terry:2026nnj} can also be investigated in coherent exclusive reactions. We address~\cite{Martinez-Fernandez:2026web,Martinez-Fernandez:2026zog} the case of the deeply bound helium-4 nucleus and demonstrate the feasibility of the extraction of a tomographic picture of this nucleus from precise coherent DVCS reaction.  

\section{Theoretical framework}

\begin{figure}[!ht] 
  \centering
  \includegraphics[width=0.8\textwidth]{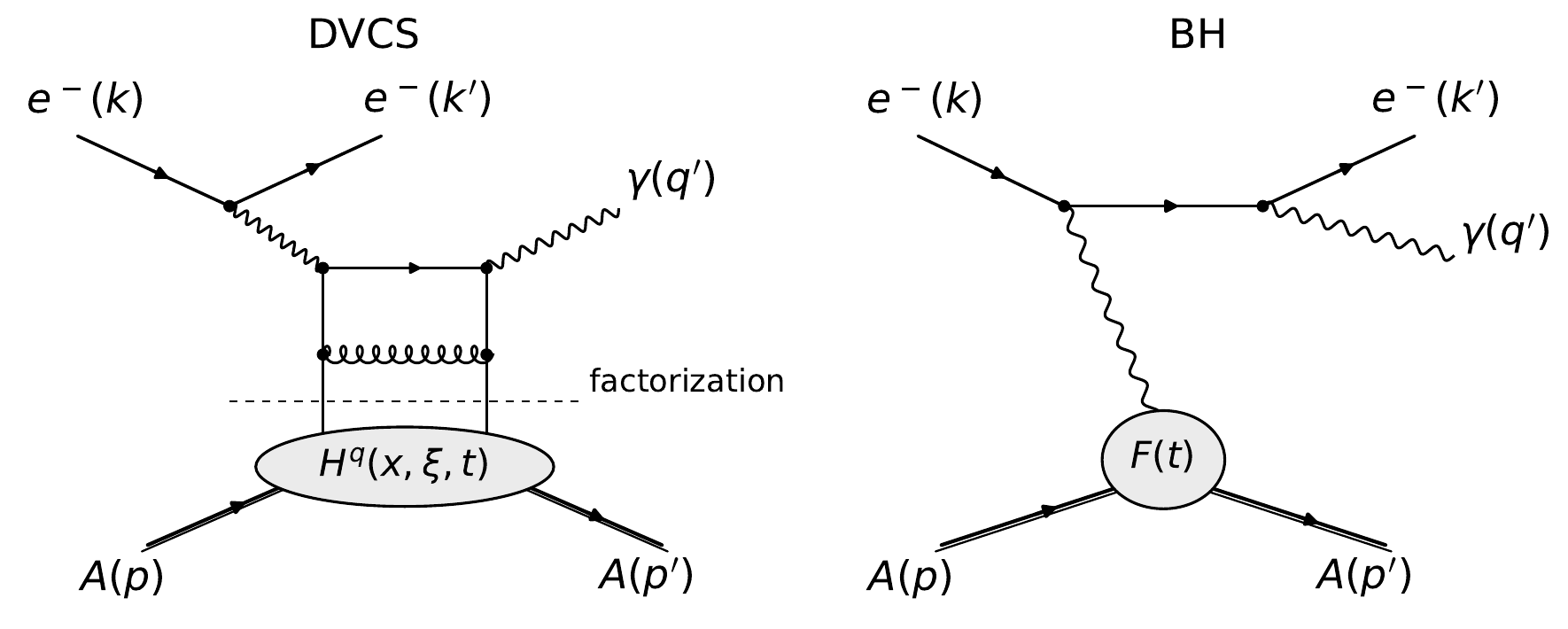}
  \caption{Examples of Feynman diagrams for  the amplitude. Left pannel : a NLO leading twist contribution; right pannel : the Bethe-Heitler contribution.}
  \label{fig::dvcs_bh}
\end{figure}

The scattering cross section for the electroproduction of a real photon takes the  form:
\begin{equation}\label{eq::cross_section}
    \frac{d^{4}\sigma^{s\lambda}}{d{x_A}\,dQ^2\,dt\,d\phi}
	= \frac{\alpha_{\rm em}^3}{8\pi} \frac{x_A y^2}{Q^4\sqrt{1+\omega^2}}
	\left|\frac{\mathcal{M}^{s\lambda}}{e^3}\right|^2 \,,
\end{equation}
where $s$ denotes the helicity of the beam and $\lambda$ denotes the polarization of the photon.  $y=(p \cdot q)/(p \cdot k)$ is the inelasticity variable, $\omega = 2x_A M/Q$, $M$ is the target mass, $x_A= Q^2/(2p \cdot q)$ the nuclear Bjorken variable, and $\phi$ is defined as the angle between the planes spanned by the leptons and by the nuclei. The amplitude is the sum of  the DVCS and Bethe-Heitler subprocesses (presented in Fig.\ref{fig::dvcs_bh}):
\begin{align}\label{eq::total_iM}
	\mathcal{M}^{s\lambda}  
    =
    \mathcal{M}^{s\lambda}_{\rm DVCS}\left(\mathcal{A}^{++}, \mathcal{A}^{+-}, \mathcal{A}^{0+}\right) 
    + \mathcal{M}^{s\lambda}_{\rm BH}(F) \,,
\end{align}
where the contributions from the DVCS and BH subprocesses are explicitly separated. Our analysis is based on the collinear QCD factorization of the DVCS helicity amplitudes into quark and gluon leading twist GPDs and coefficient functions. We include both the  next to leading order (NLO) in the strong coupling contribution~\cite{Ji:1998xh, Pire:2011st} and kinematical next to leading twist power corrections~\cite{Braun:2012bg}, neglecting genuine higher-twist contributions which  involve additional nonperturbative functions. We use full evolution equations for our model GPDs. 

\section{Modeling quark and gluon GPDs in helium-4}
\begin{figure}[!ht] 
  \centering
  \includegraphics[width=0.8\textwidth]{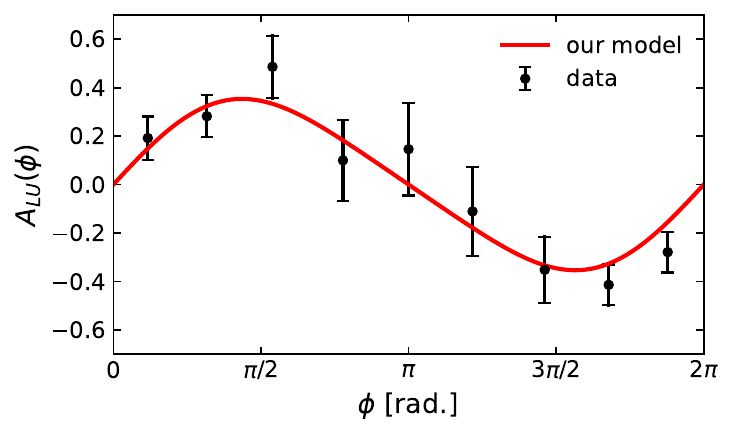}
  \caption{Fit to the beam spin asymmetry for DVCS, $A_{LU}$, measured on a helium-4 target by CLAS~\cite{CLAS:2017udk}. Only one kinematic bin is shown here: $x_B=0.172$, $t=-0.099\,\mathrm{GeV}^2$ and $Q^2=1.42 \,\mathrm{GeV}^2$.}
  \label{fig::dvcs_fit}
\end{figure}

Contrarily to Ref.~\cite{Fucini:2018gso} where a nuclear physics construction of GPDs is attempted, with the caveat of breaking the Lorentz invariance, we construct our GPD models for  $H^{i}(x,\xi,t;\, \mu^2)$, where $\mu^2$ corresponds to the energy scale, $i = \{u_{\mathrm{val}}, u_{\mathrm{sea}}, d_{\mathrm{val}}, d_{\mathrm{sea}}, s, g\}$ (isospin symmetry is assumed in our model) in the theoretically sound  double distribution framework based on measured PDFs and on experimental $t$ distribution of the electromagnetic helium-4 form factor. It reads:
\begin{equation}
H^{i}(x,\xi,t;\, \mu^2) =
\int_{-1}^{1}d\beta
\int_{-1+|\beta|}^{1-|\beta|}d\alpha \,
\delta(\beta + \xi\alpha - x)
F^{i}(\beta, \alpha, t;\, \mu^2) \,.
\end{equation}
with the double distributions $F^{i}(\beta, \alpha, t;\, \mu^2)$ expressed as products of generalized $t$-dependent PDFs, $f_{i}(\beta, t)$, and a profile function, $h_{i}(\beta, \alpha)$, which governs the buildup of the skewness effect generating the dependence on the variable $\xi$. Considering $\mu^2$-dependence implicit, we may write
\begin{equation}
F^{i}(\beta, \alpha, t) = f_{i}(\beta, t)\,h_{i}(\beta, \alpha) \,.
\label{eq:dd_master}
\end{equation}
The generalized $t-$dependent PDF is written as:
\begin{equation}
f^A_{i}(\beta, t) = f^A_{i}(\beta)\, \frac{k_{i}(|\beta|, t)}{k_{i}(|\beta|, 0)} \,.
\end{equation}
We refrain from using fully factorized dependencies in $x$ and $t$ and parametrize, for valence quarks, $i = u_{\mathrm{val}}$, 
\begin{align}
& k_{i}(|\beta|, t) = \left(\frac{1}{1-p_0(1-|\beta|)^2t}\right)^{p_1}
\prod_{j=1}^{n}\left(\frac{|p_{2,j} + t|}{|p_{2,j} + t| - p_{3,j}(1-|\beta|)^2t}\right)^{p_{4,j}} 
\,.
\label{eq:tDep}
\end{align}
This Ansatz consists of two distinct components. The first, a dipole-like Ansatz, captures the main trend in $t$ for a given $x$. The second, represented by the product, accounts for $n$ diffractive minima occurring at $-t=\{p_{2, 0}, p_{2, 1}, \ldots\}$. The free parameters of Eq.~\eqref{eq:tDep} are fitted to the helium-4 elastic form factor data, utilizing the relation
\begin{equation}
F(t) = \frac{1}{Z} \int_{-1}^{1}dx \left(\frac{2}{3}f_{u_{\mathrm{val}}}^A(x, t) + \frac{-1}{3}f_{d_{\mathrm{val}}}^A(x, t)\right) \,.
\label{eq:eff}
\end{equation}
For sea quarks and gluons,  
$i = \{u_{\mathrm{sea}}, s, g\}$, we use a different Ansatz, namely 
\begin{align}
\displaystyle{
k_{i}(|\beta|, t) = 
\exp(p(1-|\beta|^2)t)
}\,.
\label{eq:tDepSea}
\end{align}
The slope is fitted to CLAS data~\cite{CLAS:2017udk},  as illustrated for
a single kinematic bin in Fig.~\ref{fig::dvcs_fit}, and found to be $p=22.0^{+5.5}_{-2.1}~\mathrm{GeV}^{-2}$ for the best fit. 

\section{Results and conclusion}
\begin{figure}[!ht] 
  \centering
  \includegraphics[width=0.7\textwidth]{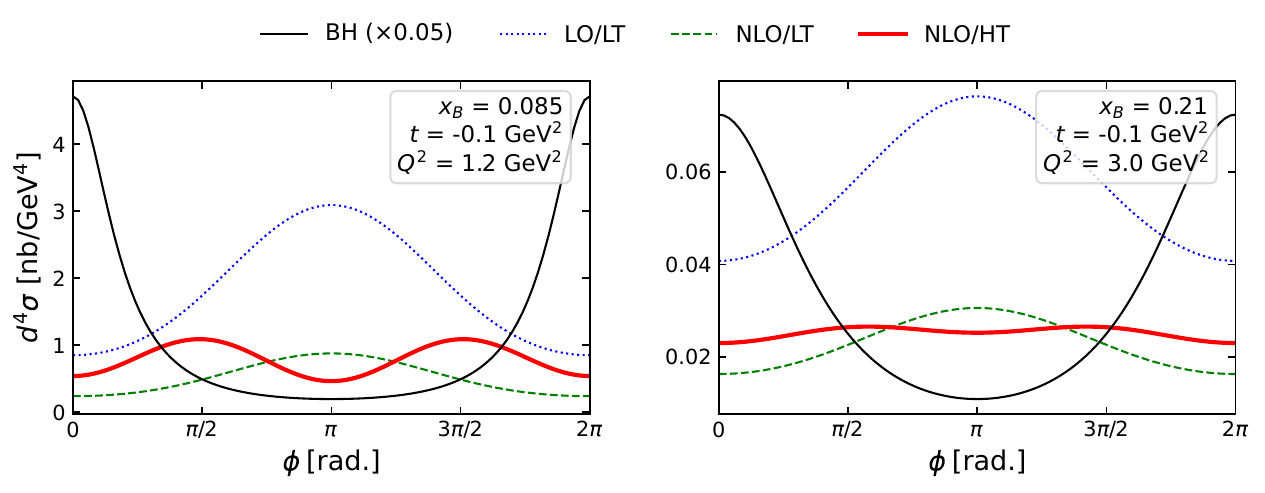}
  \includegraphics[width=0.7\textwidth]{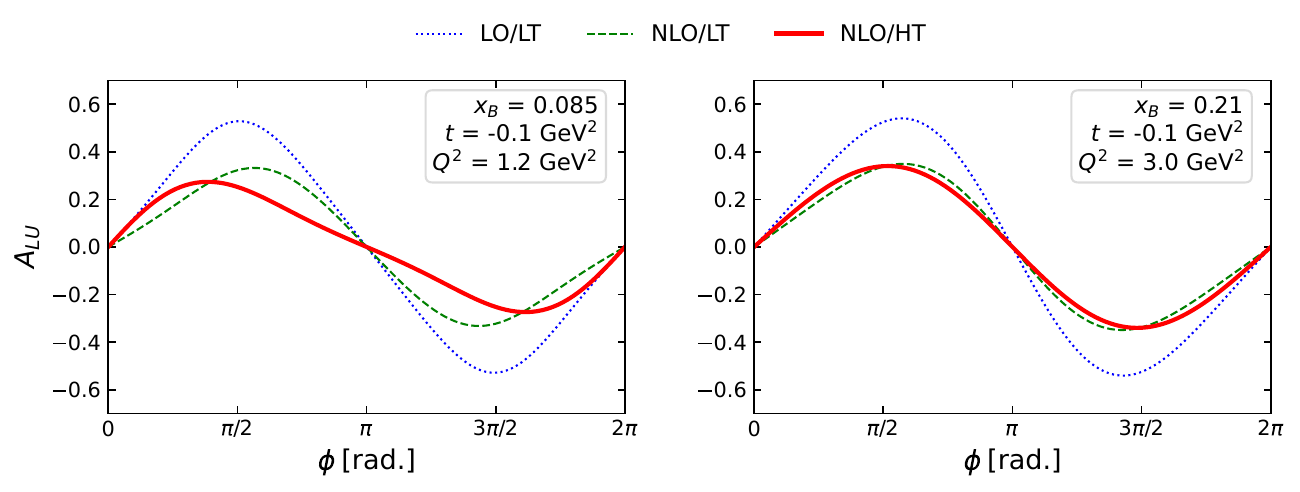}  
  \caption{Predictions for JLab12: differential cross-sections (upper row) and $A_{LU}$ asymmetries (lower row) for $Q^2 = 1.2\,\mathrm{GeV}^2$ (left) and $Q^2 = 3\,\mathrm{GeV}^2$ (right), at $y=0.75$, $t = -0.1\,\mathrm{GeV}^2$, and an electron beam energy of $10.06\,\mathrm{GeV}$.}
  \label{fig:jlab12}
\end{figure}

Using this GPD with the fitted parameters enbles us to predict cross sections and beam spin asymmetry for the near future experiments at JLab energy, as well as for EIC.
 Fig.~\ref{fig:jlab12} presents our predictions for the $A_{LU}$ asymmetries and unpolarized cross sections at typical JLab12 kinematic settings and $Q^2=1.2\,\mathrm{GeV}^2$ and $3\,\mathrm{GeV}^2$, at fixed $y$. While the Bethe-Heitler contributions are dominant in the cross-sections, the importance of NLO and kinematical higher twist contributions is clearly demonstrated by these results. This demonstrates the importance of not neglecting their effects to conduct a meaningful phenomenological extraction of GPDs , and hence to obtain a reliable tomographic image of the nucleus.

\begin{figure}[!ht] 
  \centering
  \includegraphics[width=0.8\textwidth]{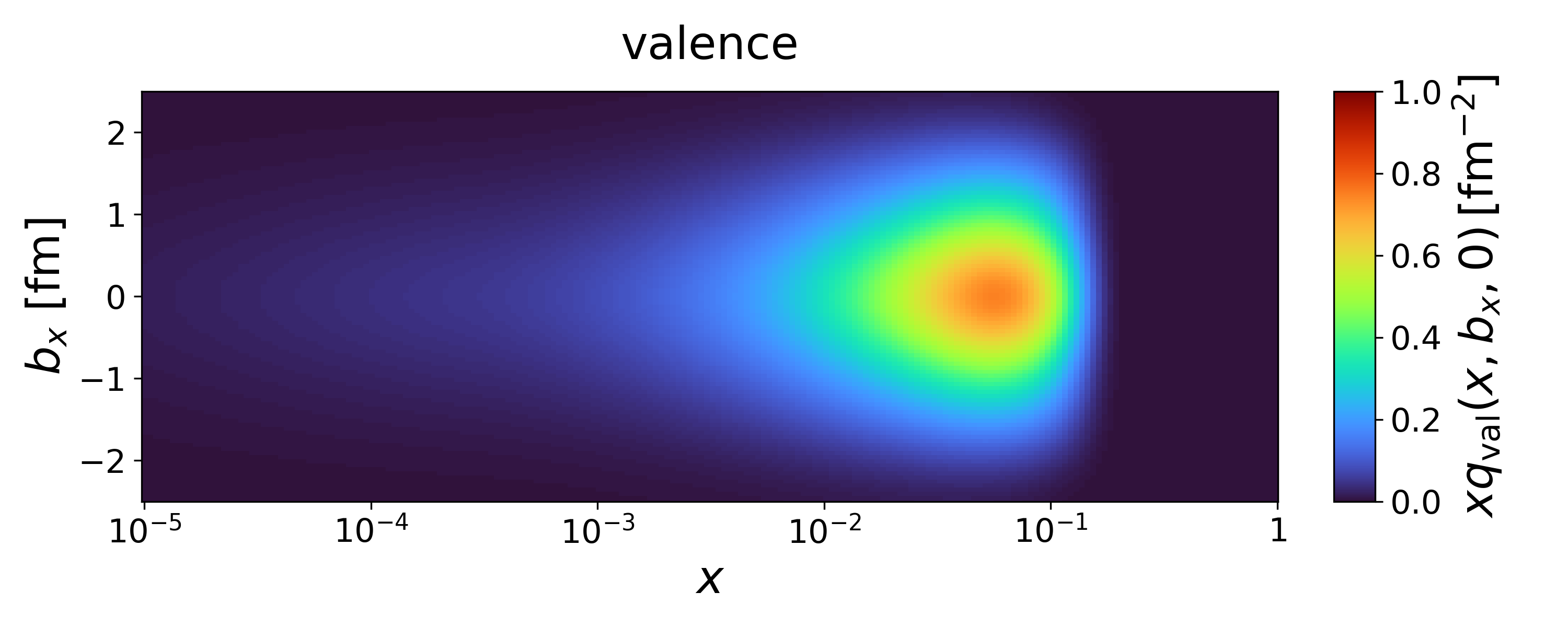}
  \includegraphics[width=0.8\textwidth]{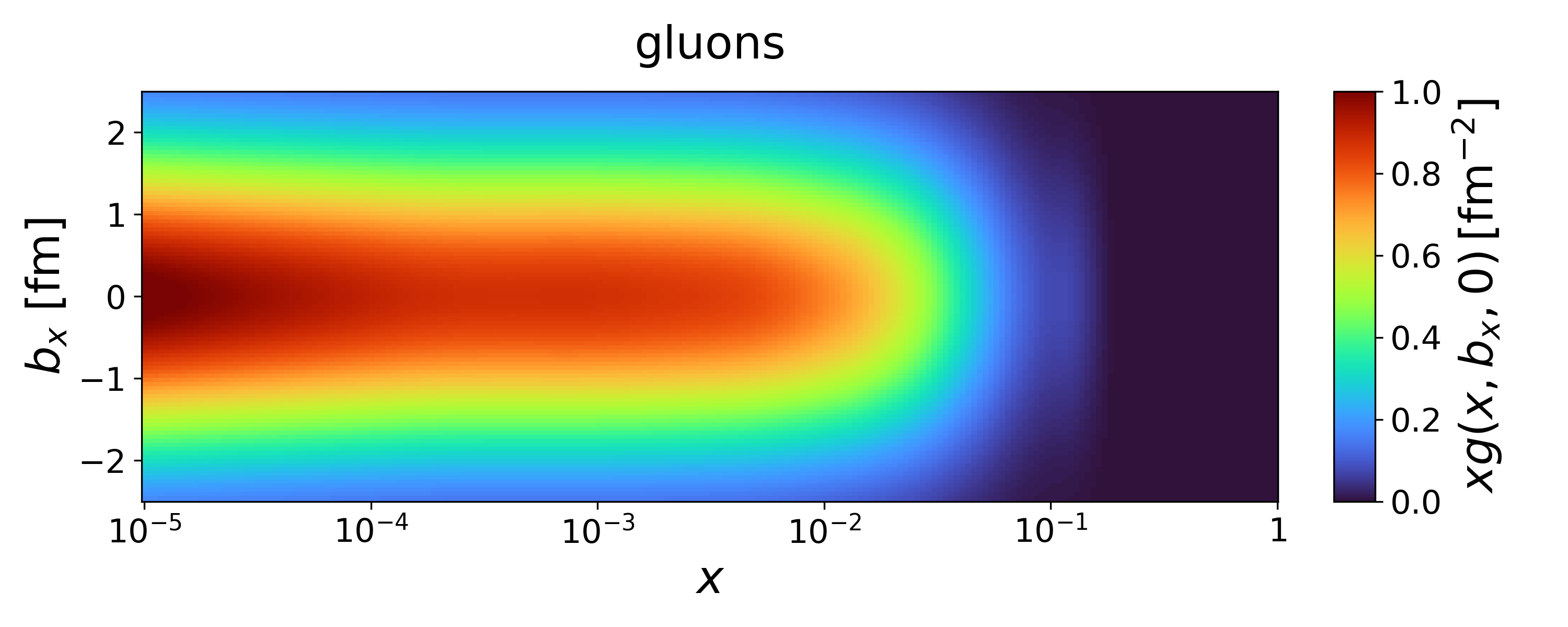}
  \caption{Spatial distributions of valence quarks, $x\,q(x, b_x, b_y=0)$, and gluons, $x\,g(x, b_x, b_y=0)$, in helium-4 nuclei at the scale $\mu^2 = 2\,\mathrm{GeV}^2$.}
  \label{fig::nt}
\end{figure}
In Fig.~\ref{fig::nt}, we present the first tomographic view of valence quarks and gluons in the helium-4 nucleus, revealing their distinct transverse spatial distributions.

In conclusion, let us stress that further studies of complementary exclusive reactions will be important for developing a more complete tomographic picture of quarks and gluons in light nuclei such as the deuteron, helium-3, and helium-4. This represents an important step toward understanding how the spatial and momentum structure of quarks and gluons is modified in a complex many-nucleon environment.

\paragraph*{Acknowledgements.}

This research was funded in whole or in part by the National Science Centre, Poland (grant IMPRESS-U No.~2024/06/Y/ST2/00155 and grant SONATA BIS-15 No.~2025/58/E/ST2/00045). For the purpose of open access, the authors have applied a CC-BY copyright licence to any Author Accepted Manuscript (AAM) version arising from this submission. The research of V.M.-F. was funded in part by l’Agence Nationale de la Recherche (ANR), project ANR-23-CE31-0019. 
\bibliographystyle{unsrt}
\bibliography{bibliography}

\end{document}